\documentclass[aps,prl,twocolumn,showpacs,superscriptaddress]{revtex4}
\usepackage{amsmath}
\usepackage{amssymb}
\usepackage{graphicx}
\usepackage{exscale}
\begin{document}


\title{Counterion density profiles at charged flexible  membranes}


\author{Christian C. Fleck}
\affiliation{Fachbereich Physik, Universit\"at Konstanz, Universit\"atsstrasse 10, 78457 Konstanz, Germany}
\author{Roland R. Netz}
\affiliation{Sektion Physik, LMU Munich, Theresienstrasse 37, 80333 Munich, Germany}

\date{\today}

\begin{abstract}
Counterion distributions at charged soft  membranes are studied using 
perturbative analytical and  simulation methods
in both  weak coupling (mean-field or Poisson-Boltzmann)  
and strong coupling limits. 
The softer the membrane, the more smeared out the counterion density profile
becomes and counterions pentrate through the mean-membrane surface location,
in agreement with anomalous scattering results.
Membrane-charge repulsion leads to a short-scale roughening  of the membrane.
\end{abstract}


\maketitle

The study of charged colloids and biopolymers  faces a fundamental problem: 
In theoretical investigations, the central object which is primarily
computed  is the charge density distribution in the electrolyte 
solution adjacent to the charged body \cite{book}.
Experimentally measurable observables are typically derived from
this charge distribution. For example, the force between charged particles
follows from the ion density at the particle surfaces via the contact-value
theorem. Likewise, the surface tension and surface potential are obtained
as weighted integrals over the ion distributions. 
It has proven difficult to measure the counterion distribution at a
charged surface directly because of the small scattering intensity. 
Notable exceptions are neutron scattering contrast variation with deuterated
and protonated organic counterions \cite{scatter1}
and local fluorescence studies on Zinc-ion distributions using X-ray standing 
waves \cite{scatter2}. 
Clearly, direct comparison between theoretical and experimental 
ion distributions (rather than derived quantities) is desirable as it
provides important hints how to improve theoretical modeling.

In a landmark paper the problem of low scattering  intensity 
 was overcome
by anomalous X-Ray scattering on stacks of highly charged bilayer
membranes \cite{rich}. Anomalous scattering techniques allow 
to sensitively discriminate counterion scattering from the background, and
a multilayer consisting of thousands of  charged layers 
gives rise to substantial scattering intensity. Since then,
similar  techniques have been applied to charged 
biopolymers \cite{Doniach,Wong} and to oriented charged bilayer stacks,
where the problem of powder-averaging is avoided \cite{Tim}. 

However, scattering on soft bio-materials brings in a new
complication, not  considered theoretically so far:
soft membranes and biopolymers 
fluctuate in shape, and thus perturb the 
counterion density profile. Comparison with standard theories
for rigid charged objects of simple geometric shape becomes 
impossible. Here we fill this gap by considering the
counterion-density profile close to a planar charged membrane
which exhibits shape fluctuations governed by  bending rigidity.
As main result, we derive for a relatively stiff membrane 
closed-form expressions for the counterion density  profile 
in the asymptotic low and high-charge  limits
which compare favorably with our simulation results.
These parametric profiles, 
which exhibit a crucial dependence on the membrane stiffness,
will facilitate the
analysis of scattering results since they allow for a data fit
with only a very few physical parameters.
In previous experiments, a puzzling ion penetration into the lipid region
was detected but interpreted as an artifact \cite{rich}. 
We  show that ion 
penetration indeed occurs and is due to the correlated ion-membrane
spatial fluctuations.
The electrostatic coupling between membrane charges  and counterions
not only modifies the counterion density profile
but also renormalizes the membrane roughness:
the short-scale bending rigidity 
is reduced,  charged membranes become locally softer.

\begin{figure}
\includegraphics[width=0.4\textwidth]{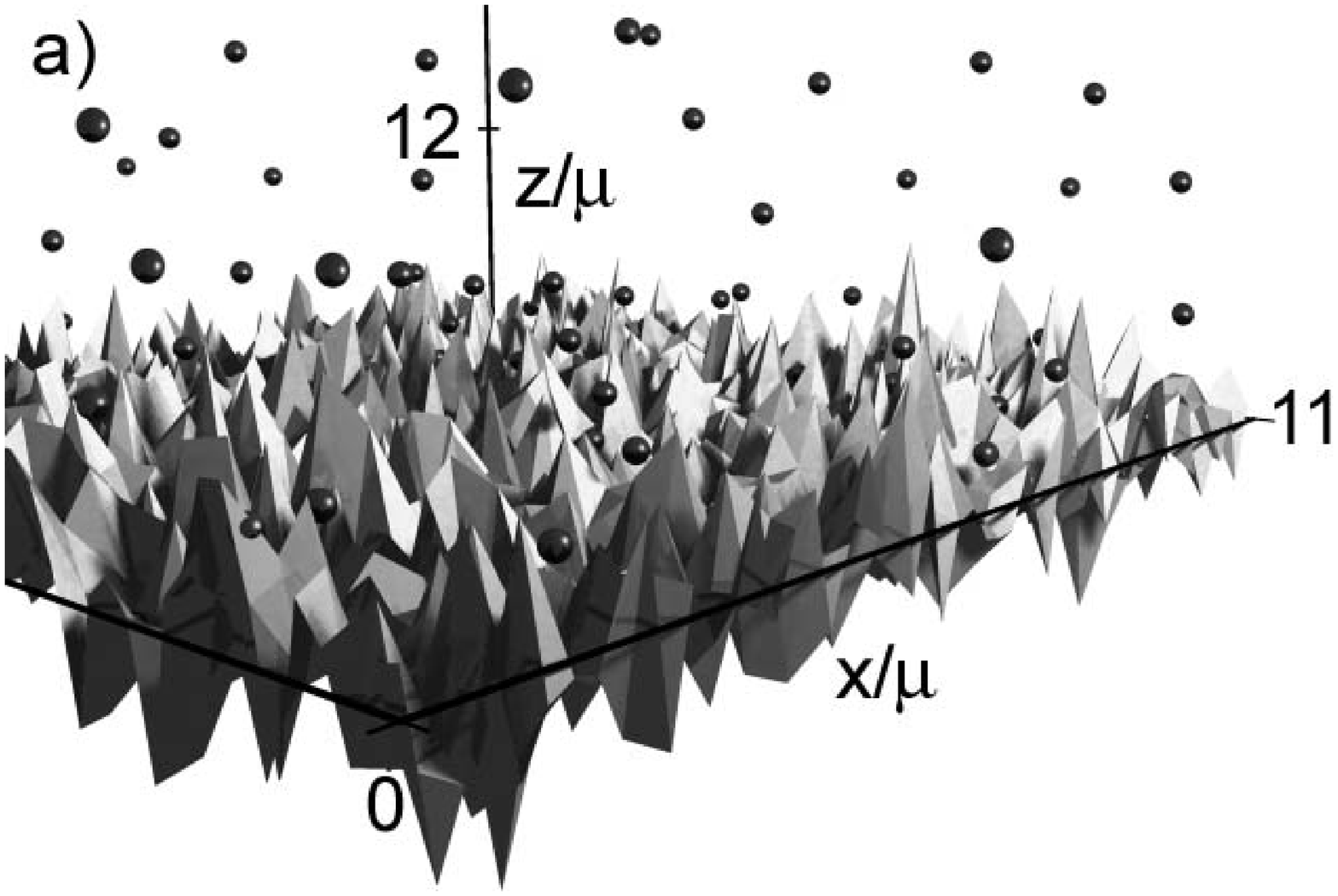}
\includegraphics[width=0.4\textwidth]{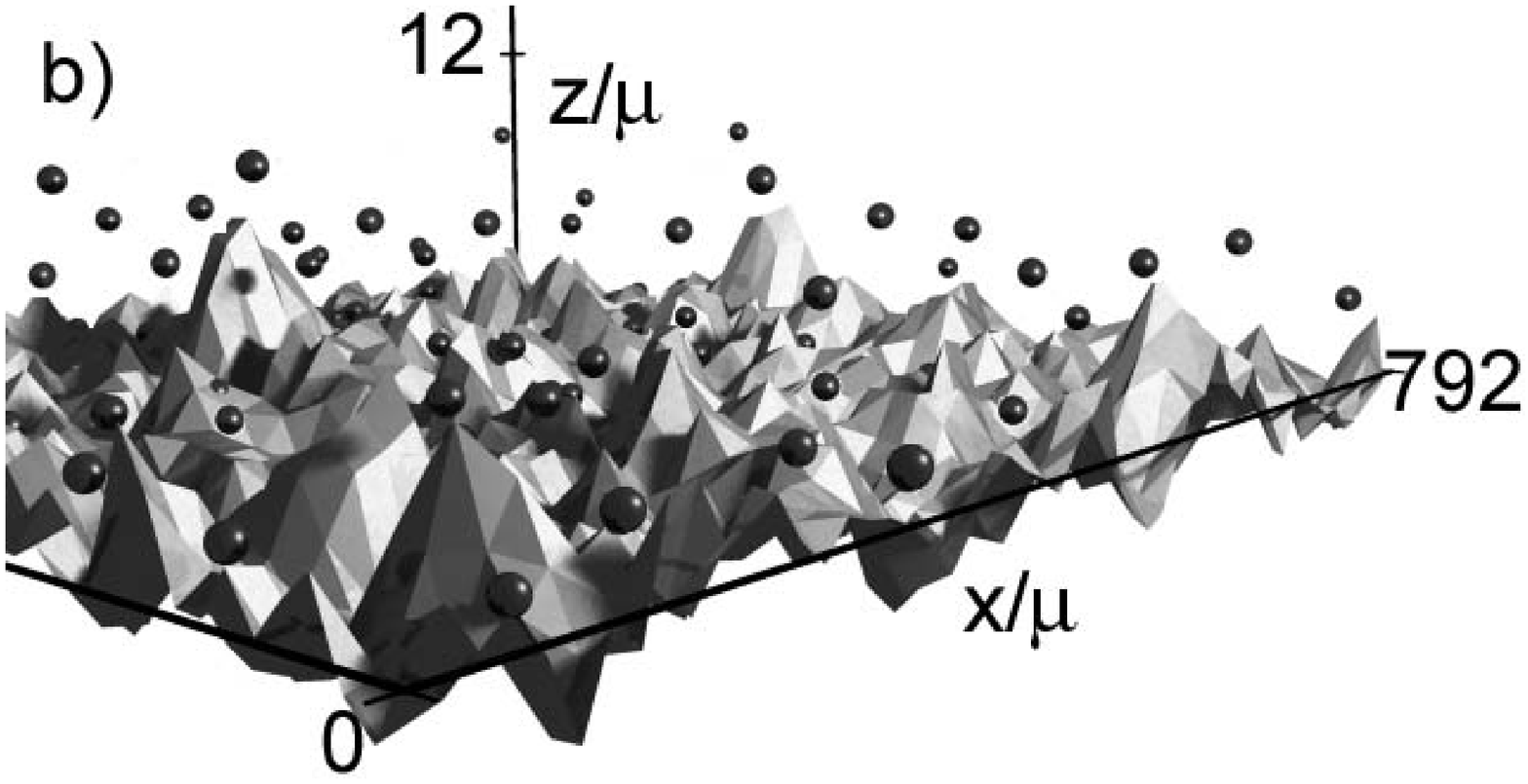}
\caption{\label{fig_snapshot} Simulation snapshots for a) 
$\Xi=0.2$, $\xi^0_{\bot}/\mu=0.80$, $K_0=0.07$, $\tilde{g}=0.57$, $\tilde{a}=0.18$, $\tilde{d}=2.2$
and b) $\Xi=1000$, $\xi^0_{\bot}/\mu=0.38$, $K_0=174$, $\tilde{g}=0.006$, 
$\tilde{a}=13.21$, $\tilde{d}=160$.
The simulations are done using $N=100$ counter-ions and 
$M=100$ membrane-ions on a $N_L =60 \times 60 $ membrane lattice.}
\vspace{-0.4cm}
\end{figure}

The Hamiltonian $H=H_m+H_e$ of the membrane-counterion system 
consists of the  elastic membrane part $H_m$ and the electrostatic part $H_e$.
We  discretize the membrane shape on a two-dimensional $N_L\times N_L$ square lattice
with lattice constant $a$
and rescale all lengths by the Gouy-Chapman length $\mu=1/2\pi q\ell_B\sigma_m$ 
according to $\mathbf{r}=\mu\tilde{\mathbf{r}}$, where $\sigma_m=QM/ N_L^2 a^2$ 
is the projected charge density of the membrane
and $\ell_B=e^2/4\pi\varepsilon_0\varepsilon k_BT$ is the Bjerrum length ($e$ is the elementary charge, $\varepsilon$ the dielectric constant). 
Parametrizing  the membrane shape by the
height function $h({\bf x})$, the elastic membrane energy in harmonic approximation 
reads in units of $k_B T$ \cite{Lip}:
\begin{eqnarray}
\label{eq_hamiltonian_mechanical_1}
H_m[\tilde{h}]&=&\frac{K_0}{2}\int
 d^2\tilde{x}\,\left(\Delta \tilde{h}(\tilde{\mathbf{x}})\right)^2 + \frac{\tilde{g}}{2}\int
d^2 \tilde{x}\,\tilde{h}^2(\tilde{\mathbf{x}}),
\end{eqnarray}
where $\Delta$ is the Laplace operator, $K_0$ is the bare bending rigidity 
and $\tilde{g}=g\mu^4$ is the rescaled strength of the harmonic potential. 
The electrostatic energy 
accounts for the interaction of $N$ counter-ions of valence $q$ and 
$M$ membrane charges of valence $Q$, related  by the electroneutrality condition $QM=qN$,
\begin{eqnarray}
\label{eq_hamiltonian_electrostatic_1}
H_e&=&\sum_{i=1}^{N-1}\sum_{j=i+1}^{N}\frac{\Xi}{\left |\tilde{\mathbf{r}}_i-\tilde{\mathbf{r}}_j\right|}-\nonumber\\
&&\sum_{i=1}^{N}\sum^M_{k=1}\frac{Q/q\Xi}{\left|\tilde{\mathbf{r}}_i-\tilde{\mathbf{R}}_{k}\right|}+\sum^{M-1}_{k}\sum^M_{l=k+1}\frac{(Q/q)^2\Xi}{\left|\tilde{\mathbf{R}}_{k}-\tilde{\mathbf{R}}_{l}\right|}
\end{eqnarray}
where $\Xi=2\pi q^3\ell_B^2\sigma_m$ denotes the coupling parameter \cite{moreira02a}. 
The rescaled position  of the  $i$th counterion  is $\tilde{\mathbf{r}}_i$ while
the $k$-th membrane-ion is located at 
$\tilde{\mathbf{R}}_{k}=(\tilde{\mathbf{x}}_{k},\tilde{h}(\tilde{\mathbf{x}}_{k})-\tilde{d})$ 
where the membrane charges are displaced by $\tilde{d}=2\tilde{a}N_L M^{-1/2}$
beneath the membrane surface which is impenetrable to the point-like counterions.
This way we can largely 
neglect charge-discreteness effects \cite{moreira02b}
and concentrate on shape-fluctuation effects.
In most of our simulations the membrane ions are mobile and move 
freely on the membrane lattice, with a packing fraction $\zeta=M/N^2_L$.
For the long-ranged electrostatic interactions we employ laterally periodic boundary conditions 
using  Lekner-Sperb methods \cite{moreira02a}. To minimize discretization and finite-size effects,
the number of lattice sites $N_L$  and the rescaled strength of the harmonic potential
$\tilde{g}$ are chosen such that the lateral height-height correlation length 
of the membrane $\xi^0_{\parallel}$ 
obeys the inequality: $\tilde{a}<\tilde{\xi}^0_{\|}=(4K_0/\tilde{g})^{1/4} \ll N_L\tilde{a}$\cite{Lip}.
Simulations are run for typically $10^6$ Monte Carlo steps
using 100 counter-ions and 100 membrane ions. 
 In Fig.\ref{fig_snapshot} we show two simulation snapshots.
 The counter-ions form in the weak coupling limit ($\Xi=0.2$, Fig.\ref{fig_snapshot}.a) a diffuse dense
 cloud while in the strong coupling limit ($\Xi=1000$, Fig.\ref{fig_snapshot}.b, 
 note the anisotropic rescaling) 
 the lateral ion-ion distances are large compared to the mean separation from the membrane.
 Pronounced correlations between membrane shape fluctuations and counterion positions
 are observed in both snapshots.

\begin{figure}
\includegraphics[width=0.4\textwidth]{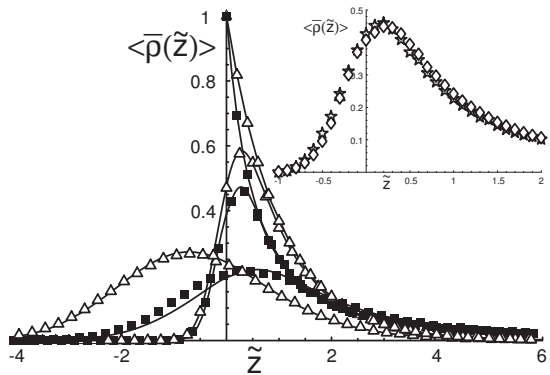}
\caption{\label{fig_pb_counter} Rescaled counterion density $\left\langle\bar{\rho}(\tilde{z})\right\rangle=\left\langle\rho(\tilde{z})\right\rangle/2\pi\ell_B\sigma_m^2$ as a function of the rescaled distance $\tilde{z}=z/\mu$ from Monte Carlo simulations
(data points) and asymptotic theory (solid lines).
In the weak coupling limit ($\Xi=0.2$, solid squares), the membrane roughness is $\tilde{\xi}^0_{\bot}=1.211\,,0.3184\,,0$ and $\tilde{\xi}^0_{\|}=0.2483\,,0.2933\,,\infty$ 
from bottom to top. In the 
strong coupling limit ($\Xi=1000$, open triangles) 
we have $\tilde{\xi}^0_{\bot}=1.211\,,0.3184\,,0$
and $\tilde{\xi}^0_{\|}=17.2475\,,20.7458\,,\infty$ from bottom to top. Numerical
errors  are smaller then the symbol sizes. 
In all cases the membrane-ions are mobile and the packing fraction is $\zeta=0.028$. 
The inset compares  profiles for $\Xi=0.2$, $\tilde{\xi}^0_{\bot}=0.3184$ for 
 $\zeta=0.028$ (diamonds) and $\zeta=0.25$ (circles) for mobile membrane ions and
results for $\Xi=0.2$, $\tilde{\xi}^0_{\bot}=1.211$, $\zeta=0.028$ for mobile (squares) and
fixed (stars) membrane ions and $\Xi=1000$, $\tilde{\xi}^0_{\bot}=1.211$, $\zeta=0.028$ 
for mobile (triangle) and  fixed (crosses) membrane ions. }
\vspace{-0.5cm}
\end{figure}

The qualitatively different ionic structures at low/high coupling strength are reflected
by fundamentally different analytic approaches in these two limits:
Starting point is the  exact expression for the partition function
\begin{eqnarray}
\label{eq_partion_function}
\mathcal{Z}=\int\mathcal{D}\tilde{h}\frac{1}{N!}
\prod_{i=1}^{N} \int {d \tilde{\bf x}_i}  \int_{\tilde{h}(\tilde{\bf x}_i)}^\infty d \tilde{z}_i
 e^{-H}\,.
\end{eqnarray}
 By performing a Hubbard-Stratonovich transformation and a transformation to the grand-canonical ensemble, we arrive at the partition function \cite{henri1}:
\begin{eqnarray}
\label{eq_partion_function_1}
\mathcal{Z} \simeq  \int\mathcal{D}\tilde{h}
\mathcal{D}\phi e^{-H_m[\tilde{h}]-H_\phi[\tilde{h},\phi,\pi]/\Xi}
\end{eqnarray}
The field $i\phi$ is the fluctuating electrostatic potential \cite{henri1}.
The  electrostatic action reads
\begin{eqnarray}
\label{eq_hamiltonian_electrostatic_2}
H_\phi [\tilde{h},\phi,\pi]\!\!&=&\!\!\frac{1}{8\pi}\!\int \!d\tilde{\mathbf{r}}\,\left(\nabla\phi(\tilde{\mathbf{r}})\right)^2\!-\!\frac{i}{2\pi}\!\int \!d\tilde{\mathbf{r}}\,\delta (\tilde{z}-\tilde{h}(\tilde{\mathbf{x}}))\phi(\tilde{\mathbf{r}})\nonumber\\
&&-\frac{\Lambda}{2\pi}\!\int  \!d\tilde{\mathbf{r}}\,e^{\pi(\tilde{\mathbf{r}})-i\phi(\tilde{\mathbf{r}})}
 \theta(\tilde{z}-\tilde{h}(\tilde{\mathbf{x}}))
\end{eqnarray}
where $\theta(z)=1$ for $z>0$ and zero otherwise.
The expectation value of the  counter-ion density 
is calculated by the help of the generating field  $\pi(\mathbf{r})$ according to 
$\langle\bar{\rho}(\tilde{\mathbf{r}})\rangle=2\pi\Xi\delta  \ln
\mathcal{Z}/\delta\pi(\tilde{\mathbf{r}})\mu^3$
and reads
\begin{eqnarray}
\label{eq_cp_1}
\left\langle\bar{\rho}(\tilde{\mathbf{r}})\right\rangle=\frac{\left\langle\rho(\tilde{\mathbf{r}})\right\rangle}{2\pi\ell_B\sigma_m^2}=\Lambda\left\langle \theta(\tilde{z}-\tilde{h}(\tilde{\mathbf{x}}))e^{-i\phi(\tilde{\mathbf{r}})}\right\rangle\,.
\end{eqnarray}
The dimensionless fugacity $\Lambda$
is determined by the normalization condition of the counterion distribution $\int d\mathbf{r}\,\left\langle\rho(\mathbf{r})\right\rangle=N$, which is in rescaled units equivalent to $
\Lambda\int d\tilde{\mathbf{r}}\,\left\langle\theta(\tilde{z}-\tilde{h}(\tilde{\mathbf{\mathbf{x}}}))e^{-i\phi(\tilde{\mathbf{r}})}\right\rangle=1$.
The partition function Eq.(\ref{eq_partion_function_1}) is intractable.
In the weak coupling limit, $\Xi\to 0$,  fluctuations  of the field $\phi$
around the saddle point value are small and gaussian variational methods become 
accurate \cite{henri2}. The variational Gibbs free energy reads:
\begin{eqnarray}
\label{eq_var_free_energy}
F_v&=&F_0+\left\langle H_\phi[\tilde{h},\phi,\pi]/\Xi+H_m[\tilde{h}]-H_0[\tilde{h},\phi]\right\rangle_0
\end{eqnarray}
Here $\langle\cdots\rangle_0$ is an average with the variational hamiltonian 
$H_0$ and $F_0$ is the corresponding free energy. The most general Gaussian
variational hamiltonian $H_0$ is
\begin{eqnarray}
H_0[\tilde{h},\phi]&=&\frac{1}{2}\int d\tilde{\mathbf{x}} d\tilde{\mathbf{x}}\prime\, \tilde{h}(\tilde{\mathbf{\mathbf{x}}})K^{-1}(\tilde{\mathbf{x}},\tilde{\mathbf{x}}\prime)\tilde{h}(\tilde{\mathbf{x}}\prime)\nonumber\\
&&+\frac{1}{2}\int d\tilde{\mathbf{r}}d\tilde{\mathbf{r}}\prime\,\Omega(\tilde{\mathbf{r}})v^{-1}(\tilde{\mathbf{r}},\tilde{\mathbf{r}}\prime)\Omega(\tilde{\mathbf{r}}\prime)\,,
\end{eqnarray} 
where the field $\Omega$ is defined by
$\Omega(\tilde{\mathbf{r}}):=\phi(\tilde{\mathbf{r}})-\phi_0(\tilde{\mathbf{r}})+
i\int d\tilde{\mathbf{x}} \prime d\tilde{\mathbf{x}}\prime\prime \,P(\tilde{\mathbf{r}};\tilde{\mathbf{x}}\prime)K^{-1}
(\tilde{\mathbf{x}}\prime,\tilde{\mathbf{x}}\prime\prime)
\tilde{h}(\mathbf{\tilde{\mathbf{x}}}\prime\prime)$
and  $P$ is the connected correlation function
$P(\tilde{\mathbf{r}};\tilde{\mathbf{x}}\prime)=\langle i\phi(\tilde{\mathbf{r}})\tilde{h}(\tilde{\mathbf{x}}\prime)\rangle_0^c$.
The variational parameters  are the mean potential $\phi_0$,
the coupling operator $P$, the propagator of the electrostatic field $v$ 
and the membrane propagator $K$. For $K$ 
we use the bare propagator of the uncharged membrane 
$K(\tilde{\mathbf{x}},\tilde{\mathbf{x}}\prime)=-4(\tilde{\xi}^0_{\bot})^2 
\mbox{kei}(\sqrt{2}|\tilde{\mathbf{x}}-\tilde{\mathbf{x}}\prime|/\tilde{\xi}^0_{\|})/\pi$,  
where the bare membrane roughness $\xi_\perp^0$ is given by 
$1/\sqrt{ 64 K_0\tilde{g}}=(\tilde{\xi}^0_{\bot})^2 =\langle\tilde{h}^2(0)\rangle_0$ \cite{Lip}.
Assuming the charge propagator $v$ to be isotropic and translational invariant
(which is an approximation)
$v$  turns out to be the bare Coulomb propagator, $v({\bf r}) = 1/r$.
The remaining variational equations $\delta F_v/\delta P=\delta F_v/\delta \phi_0=0$ 
are solved perturbatively in an asymptotic small $\tilde{\xi}^0_{\bot}$ expansion, i.e.
for a relatively stiff membrane.
The solution for $P$  for $\tilde{\mathbf{x}}=\tilde{\mathbf{x}}\prime$ 
is expressed in terms of the  Meijer's $\mathcal{G}$ function 
and reads (neglecting terms of $\mathcal{O}((\tilde{\xi}^0_{\bot})^3)$):
\begin{eqnarray}
\label{eq_P}
P_{\!\bot}(\tilde{z})\!\!=\!\!\frac{-(\tilde{\xi}^0_{\bot}\!)^2}{\sqrt{2}\pi^{\frac{5}{2}}}\mbox{erf}\!\!\left[\!\frac{\tilde{z}}{\sqrt{2(\tilde{\xi}^0_{\bot}\!)^2}}\!\right]\!\!
\mathcal{G}^{5,1}_{1,5}\!\!\left(\!\!\frac{1}{64}\!\!\left(\!\frac{\tilde{z}}{\tilde{\xi}^0_{\|}}\!\right)^{\!\!\!4}\!\left |{\frac{1}{2}\atop 0,\!\frac{1}{4},\!\frac{1}{2},\!\frac{1}{2},\!\frac{3}{4}}\right.\!\!\right)\!.
\end{eqnarray}
The result for the mean potential $\phi_0$  is given by Eq.(\ref{eq_phi})
and reduces in the limit $\tilde{\xi}^0_{\bot}\to0$ to the known Gouy-Chapmann potential 
$i\phi(\tilde{z})=2\ln(1+\tilde{z})$\cite{gouy,chap}.
We defined the auxiliary function $w(\tilde{z})$ as:
$w(\tilde{z}):=\sqrt{2(\tilde{\xi}^0_{\bot})^2/\pi}\exp\{-\tilde{z}^2/2(\tilde{\xi}^0_{\bot})^2\}-\tilde{z} \; \mbox{erfc}(\tilde{z}/\sqrt{2(\tilde{\xi}^0_{\bot})^2})$.
 The  counterion density 
 is calculated according to Eq.(\ref{eq_cp_1}) and up to third order in
 $\xi_\perp$ given by Eq.(\ref{eq_pb_profile});
it reduces to the known mean-field counter-ion density 
$\langle\bar{\rho}(\tilde{z})\rangle=(1+\tilde{z})^{-2}$ 
in the case of vanishing membrane roughness $\tilde{\xi}^0_{\bot}$ \cite{gouy,chap}. 
In Fig.\ref{fig_pb_counter} we show the laterally averaged counterion density profiles 
for weak coupling  $\Xi=0.2$ obtained from MC simulation (solid squares) for several membrane 
roughnesses $\tilde{\xi}_{\bot}$. For the comparison 
with the analytical expression Eq.(\ref{eq_pb_profile}) (solid lines)
 we use the discrete membrane propagator
$K_{mn}^{-1} = 4 K_0 (\cos[2 \pi n/N_L] + \cos[2 \pi m/N_L] -2)^2/a^4 + g$ 
 and calculate the membrane roughness according to
$(\tilde{\xi}^0_{\bot})^2 = \sum_{m,n} K_{mn}$.
The lateral correlation length  follows as
$\tilde{\xi}^0_{\|}=1/(2 \tilde{\xi}^0_{\bot} \tilde{g}^{1/2} )$. 
For $\tilde{z}>\tilde{\xi}^0_{\bot}$ the counterion profile approaches the corresponding profile for a planar surface, but for $\tilde{z}<\tilde{\xi}^0_{\bot}$ we find  pronounced deviations
 from the flat surface profile. For $\tilde{\xi}^0_{\bot}=1.211$ the analytical result 
 and the simulation result disagree, showing the limitation of our 
 small $\tilde{\xi}^0_{\bot}$ expansion.
\vspace{-0.7cm}
\begin{widetext}
\vspace{-0.5cm}
\begin{eqnarray}
\label{eq_phi}
i\phi_{0}(\tilde{z})&=&\left\{\begin{array}{r@{\;:\;}l} w(\tilde{z})+
2\ln\left[1+ \tilde{z}- \tilde{z} w(\tilde{z}) /4 -
(\tilde{\xi}_{\bot}^0 /2 )^2 
\mbox{erf}\left(\frac{\tilde{z}}{\sqrt{2} \tilde{\xi}_{\bot}^0}\right)\right]
+\mathcal{O}((\tilde{\xi}_{\bot}^0)^3 ) & \tilde{z} \ge 0\\
2\tilde{z}-\tilde{z}^2+w(\tilde{z})\left(1 - \tilde{z}/2 \right)-
(\tilde{\xi}_{\bot}^0)^2 \mbox{erf}\left[\frac{\tilde{z}}{\sqrt{2}\tilde{\xi}_{\bot}^0}\right]/2
+\mathcal{O}((\tilde{\xi}_{\bot}^0)^3) & \tilde{z} < 0
\end{array}\right.
\end{eqnarray}
\begin{eqnarray}
\label{eq_pb_profile}
\left\langle\bar{\rho}(\tilde{z})\right\rangle&=&\frac{e^{-i\phi_{0}(\tilde{z})}}{2}\left\{\left(1+\mbox{erf}\left[\frac{\tilde{z}}{\sqrt{2}\tilde{\xi}^0_{\bot}}\right]
\right)\left(1 -\frac{P_{\bot}(\tilde{z})}{2}\mbox{erf}\left[\frac{\tilde{z}}{\sqrt{2}\xi_{\bot}^0}\right]\right)\right.\left.
+2P_{\bot}(\tilde{z})\frac{e^{-\frac{\tilde{z}^2}{2(\tilde{\xi}^0_{\bot})^2}}}
{\sqrt{2\pi} \tilde{\xi}^0_{\bot}}\right\}+\mathcal{O}((\tilde{\xi}_{\bot}^0)^{3})
\end{eqnarray}
\vspace{-0.3cm}
\end{widetext}
In the strong coupling limit $\Xi\to\infty$ we
expand the partition function (\ref{eq_partion_function_1}) 
in inverse powers of $\Xi$ \cite{moreira02a}. 
Starting point is the exact expression Eq.(\ref{eq_cp_1}). 
After some manipulation we find for the leading term:
\begin{eqnarray}
\label{eq_cp_sc_1}
\left\langle\bar{\rho}(\tilde{\mathbf{r}})\right\rangle\!&=&\!\frac{\Lambda e^{-\Xi v(\mathbf{0})}
}{Z}\int\!\!\mathcal{D}\tilde{h}\,\theta(\tilde{z}-\tilde{h}(\tilde{\mathbf{x}}))e^{-H_m[\tilde{h}]}\nonumber\\
&&\times e^{\frac{1}{2\pi}\int\,d\tilde{\mathbf{r}}\prime \delta\left(\tilde{z}\prime-\tilde{h}(\tilde{\mathbf{x}}\prime)\right)v(\tilde{\mathbf{r}},\tilde{\mathbf{r}}\prime)}+\mathcal{O}(\Xi^{-1}).
\end{eqnarray}
This strong coupling expansion is equivalent to a virial expansion, and hence
the leading term corresponds to  the interaction of a single counterion with a 
fluctuating charged membrane \cite{moreira02a}.
For stiff membranes we can employ a small-gradient expansion,
$\frac{1}{2\pi}\int\,d\tilde{\mathbf{r}}\prime\,\delta(\tilde{z}\prime-\tilde{h}(\tilde{\mathbf{x}}\prime)v(\tilde{\mathbf{r}}-\tilde{\mathbf{r}}\prime) \simeq
C-\tilde{z}+\int d\tilde{\mathbf{r}}\prime\,\tilde{h}(\tilde{\mathbf{x}}\prime)f_{\tilde{h}}(\tilde{\mathbf{r}},\tilde{\mathbf{r}}\prime)$, where $C$ is an unimportant constant and the function $f_{\tilde{h}}(\tilde{\mathbf{r}})$ is defined by: $
f_{\tilde{h}}(\tilde{\mathbf{r}},\tilde{\mathbf{r}}\prime):=\delta(\tilde{z}\prime\!-\!\tilde{h}(\tilde{\mathbf{x}}\prime))\times$ $\left((\tilde{z}-\tilde{z}\prime)\!-\!(\tilde{\mathbf{x}}\!-\!\tilde{\mathbf{x}}\prime)\cdot\nabla\prime\tilde{h}(\tilde{\mathbf{x}}\prime)\right)\!/2\pi\left(|\tilde{\mathbf{x}}\!-\!\tilde{\mathbf{x}}\prime|^2\!+\!(\tilde{z}\!-\!\tilde{z}\prime)^2\right)^{3/2}$. Expanding Eq.(\ref{eq_cp_sc_1}) in powers of $f_{\tilde{h}}$ gives rise to:
\begin{eqnarray}
\label{eq_sc_profile}
\left\langle\bar{\rho}(\tilde{\mathbf{r}})\right\rangle\!\!=\!\!\frac{e^{-\tilde{z}-\frac{(\tilde{\xi}^0_{\bot}\!)^2}{2}}}{2}\!\!\left(\!\!1\!+\!\mbox{erf}\!\!\left[\frac{\tilde{z}}{\sqrt{2(\tilde{\xi}^0_{\bot}\!)^2}}\right]\!\right)\!\!+\!\mathcal{O}\!\left(\!\frac{1}{\Xi},f_{\tilde{h}}\!\right)\!\! .
\end{eqnarray}
The density (\ref{eq_sc_profile}) reduces to the known SC  density $\left\langle\bar{\rho}(\tilde{z})\right\rangle=e^{-\tilde{z}}$ in the limit $\tilde{\xi}^0_{\bot}\to 0$ \cite{moreira02a}. 
We compare in Fig.\ref{fig_pb_counter} the analytically obtained counterion 
density profiles (solid lines) with the laterally averaged  densities 
obtained using MC simulations (open triangles) for $\Xi=1000$ and different $\tilde{\xi}^0_{\bot}$. The analytic approximation  reproduces the simulated 
 profiles very well. Similar to the weak coupling case, the profiles approach the corresponding strong coupling density for counter-ions at a planar surface for $\tilde{z}\gg\tilde{\xi}^0_{\bot}$, but deviate noticeable from the planar distribution for $\tilde{z}<\tilde{\xi}^0_{\bot}$. 
Comparison of mobile and immobile membrane ions gives no detectable difference
for the counterion profle (Fig.2 inset).

\begin{figure}[top]
\includegraphics[width=0.4\textwidth]{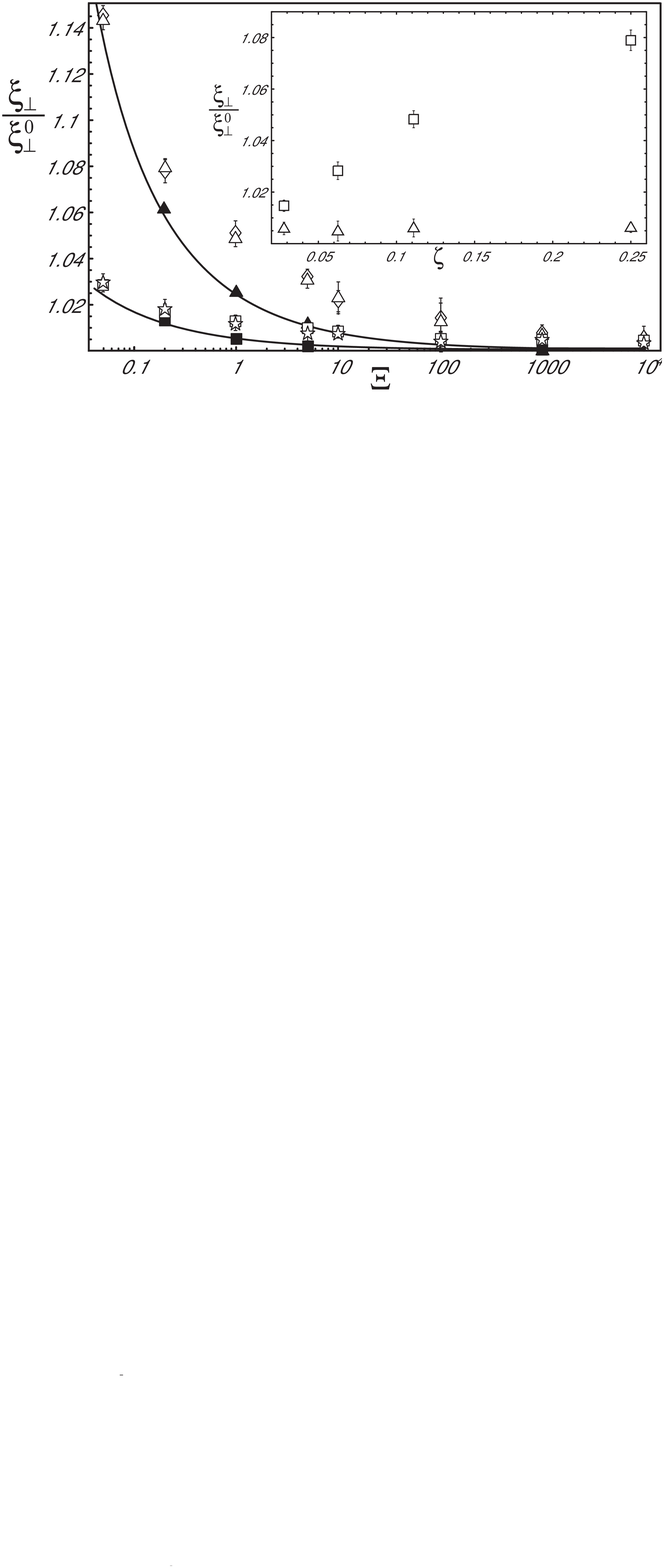}
\caption{\label{fig_roughness} Ratio of simulated and bare roughness  
$\tilde{\xi}_{\bot} / \tilde{\xi}_{\bot}^0$ as a function of $\Xi$ for
$\zeta=0.028$ and  $\tilde{\xi}_{\bot}^0=0.3184$ (open squares) and 
$\tilde{\xi}_{\bot}^0=1.2111$ (open stars), $\zeta=0.25$ and  $\tilde{\xi}_{\bot}^0=0.3184$ 
(open triangles) and $\tilde{\xi}_{\bot}^0=1.2111$ (open diamonds). 
The solid lines and solid symbols  are analytical and MC  results without counterions 
($\zeta=0.028$ lower branch, $\zeta=0.25$ upper branch). 
The inset shows the ratio $\tilde{\xi}_{\bot}/\tilde{\xi}_{\bot}^0$ as a function of the packing fraction $\zeta$ for $\Xi=0.2$ (squares) and $\Xi=1000$ (triangles),  
$\tilde{\xi}_{\bot}^0=0.3184$ in both cases.}
\vspace{-0.4 cm}
\end{figure}

In the analytics so far  we used the bare membrane roughness $\tilde{\xi}^0_{\bot}$
without modification due to electrostatics. 
In Fig.\ref{fig_roughness} we show the ratio of $\tilde{\xi}_{\bot}$, 
the membrane roughness measured in the MC simulation, and 
$\tilde{\xi}_{\bot}^0$, for the bare uncharged membrane,
 as a function of the coupling parameter $\Xi$ for two different surface fractions $\zeta$
 (open symbols). The ratio is larger than unity, i.e. charges on the membrane
 increase the roughness. 
 This short-range roughening, which allows membrane charges to increase
 their mutual distance and is thus not area-preserving,  has to be distinguished from the
 electrostatic stiffening in the long-wavelength limit which has been
 predicted on the mean-field-level \cite{helf,lek,mitch}. Local roughening
 corresponds to  protrusion degrees of freedom of single lipids. 
Yet a distinct softening mechanism, effective at intermediate wavelengths,
 is due to electrostatic  correlations effects \cite{phil,me,woon},
which is missed by standard mean-field approaches.
Experimentally, both membrane stiffening \cite{rowat} and, for highly charged membranes,
softening has been observed \cite{zemb}.
To distinguish effects due to membrane charges and counterions we calculate
via exact enumeration and 
within harmonic approximation the membrane propagator $K_{mn}$ for a 
charged discrete membrane without counterions. The  roughness
ratio from this analytical calculations is shown as a solid line, and again cross-checked
by MC simulations without counterions (filled symbols). 
The good agreement with the MC data containing counterions shows that the softening effect
is mostly due to the repulsion of charges on the membrane itself. Experimentally, this
short-scale roughening will show up in diffuse X-ray scattering data.
%
\begin{acknowledgments}
Financial support by the "International Research Training Group Soft Condensed Matter" at the University of Konstanz, Germany, is acknowledged. 
\end{acknowledgments}


\end{document}